\documentstyle[epsf]{l-aa}
\newcommand{\eqb}{\begin{eqnarray}}
\newcommand{\eqe}{\end{eqnarray}}
\newcommand{\md}{{\rm d}}
\newcommand{\pesc}{P_{\rm\scriptscriptstyle esc}}

\newcommand{\sca}{s_{\rm c}}
\begin{document}
\thesaurus{12(02.01.1; 02.19.1; 03.13.4; 09.03.2)}
\title{First order Fermi acceleration at multiple oblique shocks}
\author{U.D.J. Gieseler \and T.W. Jones}
\institute{ University of Minnesota, Department of Astronomy,
        116 Church St. S.E., Minneapolis, MN 55455, U.S.A.}
\offprints{gieseler@msi.umn.edu}
\date{Received 14 October 1999 / Accepted 2 March 2000}
\maketitle
\begin{abstract}
Numerical results for particle acceleration at multiple
oblique shocks are presented. We calculate the steady state spectral slope of 
test particles accelerated by the first order Fermi process. The results are
compared to analytical treatments, for parameters, where the diffusion
approximation does apply. Effects of injection and finite shock extend 
are included phenomenologically. We find the spectrum of accelerated 
particles to harden substantially at multiple oblique shocks and discuss
the influence of the number of shocks compared to the obliquity itself.
\keywords{Acceleration of particles -- Shock waves -- Methods: numerical
 -- Cosmic rays}
\end{abstract}
\section{Introduction}
\label{intro}
The theory of first order Fermi acceleration is commonly used to understand
non thermal particle spectra or source distributions of synchrotron
radiation in various astrophysical objects (for reviews see 
Drury~\cite{Drur83}; Blandford \& Eichler~\cite{BlEi87}; 
Kirk et al.~\cite{KiMePr94}). In the diffusion 
approximation the steady-state spectral index of accelerated test particles 
is solely a function of the compression ratio of the shocked plasma flow. 
For strong shocks, the compression ratio is given by
$r=\rho'/\rho=4$, where $\rho'$ and $\rho$ are the downstream and upstream
densities respectively. This leads to a phase space distribution
of accelerated particles $f(p)\propto p^{-s}$ with the canonical
spectral index $s=\sca:=3r/(r-1)=4$.
The synchrotron emission of such an (e.g.) electron distribution 
as a function of frequency is given by $\epsilon(\nu)\propto \nu^{-\alpha}$,
where $\alpha=(s-3)/2$. Unless the injection into the acceleration 
process is very effective, (leading to a strongly modified shock structure),
this is the hardest spectrum which can be produced through first order 
Fermi acceleration by a single shock
for which the diffusion approximation is valid.
In complex large scale structures like the galaxy or active galactic 
nuclei (AGN)
particles may encounter several shocks. If the transport time between 
shocks is larger than the acceleration time at a single
shock, the problem can be described by subsequent shocks, where (in the case
of planar geometry) the downstream distribution is transported
(and decompressed) upstream of the next shock.

It is well known from the theory of diffusive shock acceleration, that 
the supply of an upstream power law distribution with the 
canonical spectral index leads to an amplification of the distribution,
and the spectral index is not changed.
However, the increased number of high energy particles is accompanied by a 
decrease of the number density at the low energy cutoff, leading to a 
flattening of the distribution at intermediate energies. 
This can be seen by calculating the
spectrum as it is processed trough a number of shocks, suppressing new
injection, and considering adiabatic decompression, as shown by
Melrose \& Pope~(\cite{MePo93}).
At sufficiently high energies, a power law with the canonical spectral index 
is always revealed. This applies for multiple 
identical shocks without losses, which are subject of this 
work.\protect\footnote{For a consideration 
of synchrotron losses in the framework of multiple shocks, 
see e.g. Marcowith \& Kirk~(\protect\cite{MaKi99}) and the
references therein.}

In the limit of an infinite number of subsequent shocks with injection at
each shock, the flattening of
the spectrum (compared to a single shock) extends even to the highest energy
particles, with a momentum dependence of $f(p)\propto p^{-3}$
(White~\cite{Whit85}; Achterberg~\cite{Acht90}).

Since the modification of the particle spectrum evolves
from the low energy part of the spectrum due to further  
acceleration and adiabatic decompression, effects of injection can be very 
important. Including a theory of the injection process is
well beyond the scope of this work. However, given a momentum $p_0$
at which particles are injected, we assume, that this is the momentum
that divides particles which are able to diffuse across the shock from
the thermal pool. Adiabatic decompression can shift the momentum of
some particles to $p<p_0$. These particles are considered to be
're-thermalised' and no longer take part in the acceleration process and
the resulting spectrum.

In order to compare our numerical work with analytical treatments we first
consider subsequent identical shocks with a quasi-parallel magnetic field,
leading to diffusive acceleration. The relative orientation of magnetic 
field and shock normal is, however, very likely to be oblique.
Then, the diffusion approximation may no longer apply.
To investigate the principal effect of a low number of multiple oblique 
shocks, we use for each shock the same inclination angle $\Phi$
between magnetic field and shock normal in the upstream rest frame.
We use a fixed escape probability $\pesc$ for particles during the
propagation between subsequent planar shocks to account for effects 
of finite shock extension.\footnote{The probability $\pesc$, which we
introduced here, should not be confused with the escape probability for each 
shock acceleration cycle.} The acceleration process 
is described by the shock-drift mechanism. We take into account that multiple 
shock encounters are possible. For highly oblique 
geometry, repeated reflections off the compressed downstream field 
are the most important acceleration process. The transport upstream and
downstream is of diffusive character, but the phase space 
distribution at the shock can be highly anisotropic, and the standard 
diffusion approximation of the acceleration problem (namely, $f(x,p,\mu)$ 
is only to second order anisotropic) does not apply.

Even if we consider shocks with different inclination angles $\Phi$,
each shock will produce locally a characteristic pitch-angle distribution, 
and spectral index. Only during the transport from one shock to the
next, the pitch-angle distribution is isotropised.
Anastasiadis \& Vlahos~(\cite{AnVl93}) used a random 
pitch angle before every electron shock interaction. This isotropisation
produces steeper spectra as opposed to pitch-angle scattering, as shown
by Naito \& Takahara~(\cite{NaTa95}).

After referring to some analytical results in Sect.~\ref{analytical},
we compare our Monte-Carlo results for multiple identical planar shocks
to analytical calculations in Sect.~\ref{diffusive}, where we also include 
effects of finite spatial extend of the shocks, and re-thermalisation.
In Sect.~\ref{oblique} we present results for
multiple oblique shocks.
\section{Analytical considerations}
\label{analytical}
In the diffusion approximation, the problem of subsequent
shocks can be solved analytically. For an upstream phase-space distribution
 $f_{\rm u,1}(p,p_0)$, the downstream distribution without additional 
injection is given by 
\eqb\label{f_d}
f_{\rm d,1}(p,p_0)= \sca\,p^{-\sca}\int\limits_{p_0}^p \md p' (p')^{\sca-1}
                     f_{\rm u,1}(p')\,.
\eqe
During transport to the next shock, due to conservation of the phase space 
volume $p^3/\rho$, we have to consider the effect of the decompression of 
the plasma on the momentum of the particles. 
Expansion of the plasma by the compression 
ratio $r$ leads to the shift of the downstream momentum $p'$ to the new
upstream momentum $p=p'/r^{1/3}$ at the next shock 
(Schneider~\cite{Schn93}). 
Applying Eq.~(\ref{f_d}) to $N$ subsequent identical shocks with adiabatic
decompression between them, we get the spectral index
downstream of the $N$th shock for a delta-function injection distribution at 
$p_0$ only at the first shock (see Melrose \& Pope~\cite{MePo93}):
\eqb\label{index}
s=\frac{3 r}{r-1}-\frac{\log(e)}{\displaystyle\frac{\log(p/p_0)}{N-1}+
                  \frac{1}{3}\ln(r)\log(e)}\equiv s_{\rm c}-\Delta \,.
\eqe
For exactly identical shocks, we have to consider 
injection at all shocks. The downstream distribution at the last shock is then
given by a sum of the distributions injected with
 $f(p,p_0)=A\cdot\delta(p-p_0)$ at each shock:
\eqb\label{inj_all}
f_{{\rm d},\scriptscriptstyle N}(p,p_0)=\sum\limits_{n=0}^{N-1} 
   \frac{A\, \sca^{n+1}}{n!\,p_0}\left(r^{\frac{n}{3}}\frac{p}{p_0}
    \right)^{-\sca}
   \left[\ln\left(r^{\frac{n}{3}}\frac{p}{p_0}\right) \right]^{n},
\eqe
where the $n$th contribution is calculated by subsequent application of 
Eq.~(\ref{f_d}), and considering decompression between the shocks, as
shown by Melrose \& Pope~(\cite{MePo93}).
The spectral index\footnote{We call the locally
defined slope of the spectrum always spectral {\em index}, even where it
is not a pure power law.} is given by
 $s=-\partial\ln[f_{{\rm d},\scriptscriptstyle N}(p,p_0)]/\partial\ln(p)$,
which can be easily calculated for $N=5$ shocks, which we consider here.
For an infinite number of shocks, the result asymptotes to
 $f_{\rm d,\infty}(p,p_0)\propto p^{-3}$ (e.g., White~\cite{Whit85}).
\section{Monte-Carlo simulations}
\label{numerical}
We present results for test-particle acceleration at
multiple quasi-parallel and oblique shocks.
We have used a Monte-Carlo code which was described by 
Gieseler et al.~(\cite{GiKiGaAc99}), and extended it to multiple shocks.
We refer to this previous work for a more detailed description of the code, and
summarize here only the main features.
A particle is described by three coordinates: the distance from the 
shock, the magnitude of the momentum $p=|\vec{p}|$, and the pitch angle 
$\mu=\cos\alpha = \vec{p}\cdot\vec{B}/(pB)$ between particle momentum
 $\vec{p}$ and magnetic field $\vec{B}$. Upstream and downstream 
the particle momentum is conserved, and the small scale irregularities
lead to pitch-angle scattering. We do not consider cross field diffusion. 
The magnetic moment $p(1-\mu^2)/B$ is always conserved in the non-relativistic
shocks with velocity $u_{\rm s}=0.1c\equiv 0.1$, which we consider here. 
When a particle crosses the shock, the momentum and pitch angle are 
transformed to the new relevant rest frame. Depending on the pitch angle
 $\mu$ and
inclination angle $\Phi$, particles can be
reflected upstream of the shock. If a particle reaches a distance of a few
diffusion length scales downstream of the shock where the density 
distribution has reached its constant downstream value, it is considered as
escaped from the single shock. Between shocks, decompression of the
plasma leads to a shift in the momentum distribution, as described in 
Sect.~\ref{analytical}.

For exactly planar shocks, every escaping particle 
will reach the next shock, and can be further accelerated.  To account 
approximately for
the more realistic situation that some particles found in the downstream 
region of the $N$th shock have been accelerated only at $n<N$ shocks,
we use the fixed escape probability $\pesc$. This gives the fraction of
particles of the downstream distribution of each shock that will not 
be further accelerated. These particles remain in the system but bypass
the following shocks and therefore contribute directly to the 
downstream distribution of the $N$th shock. Exactly planar shocks would be 
described by $\pesc=0$, whereas $\pesc=1$ reveals a spectrum produced by one
single shock. 

The distinction of particles which are able to diffuse (and for which the 
shock is a discontinuity) from the thermal particles, which are not subject
to the first order acceleration process, is based on the gyro radius, and 
therefore on the momentum of the particles. We do not include a self consistent
injection model here. Instead we inject particles with a momentum
 $p_0/m=\gamma v =10$,
for which they have an already relativistic velocity, and can be 
accelerated immediately. However, when a particle has a momentum $p<p_0$,
we consider this particle as re-thermalised and suppress further acceleration,
even though the momentum $p_0$ does not correspond to a mean thermal momentum.
\subsection{Multiple shocks with diffusive acceleration}
\label{diffusive}
\begin{figure}[t]
    \vspace{-3.2cm}
    \begin{center} 
      \epsfxsize9.0cm 
      \mbox{\epsffile{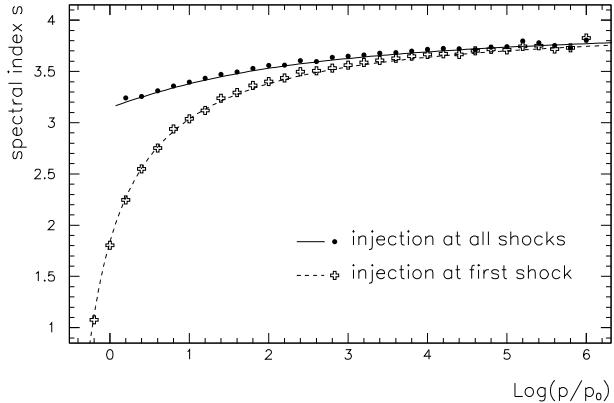}}
    \end{center}
    \vspace{-1.2cm} 
    \protect\caption{Spectral index $s$ vs. momentum $p$, downstream of 5th 
                   quasi-parallel shock. 
                   Discrete symbols show the Monte-Carlo results, whereas 
                   the lines represent analytical calculations.
                   Parameters for all shocks are $r=4$, $u_{\rm s}=0.1$,
                   $\Phi=8.07$,
                   $u_{\rm s}/\cos\Phi=0.101$. $\pesc=0$ and
                   no re-thermalisation is used.}
    \label{diffuse}
\end{figure}
In order to compare our Monte-Carlo simulations with analytical calculations,
we first consider identical planar quasi-parallel shocks. Here, {\em all}
particles are transported through the subsequent shocks with decompression 
between them. This corresponds to $\pesc =0$ for the escape probability.
We also do not include effects of re-thermalisation. Therefore we can compare
the results of the Monte-Carlo simulations directly to the analytical 
results described in Sect.~\ref{analytical}. 
To retrieve the spectral index from the momentum distribution, we use the 
linear interpolation between two neighboring momentum bins in the region 
$\log(p/p_0) < 3$. The statistical fluctuation is then 
expressed solely by the scatter of the result in this region. For higher 
momenta, where the distribution deviates only slightly from a pure power law,
we fit a power law over four equidistant momentum bins in $\log(p)$.
The result for $7.3\cdot 10^5$ particles is presented in 
Fig.~\ref{diffuse}. The open crosses represent
particles which are injected at the first shock 
with $p=p_0$.
The final distribution is measured downstream of the
5th shock before decompression. Equation~(\ref{index}) is shown by the
dashed line. The filled dots show the spectral slope in the case when 
particles are injected
at all shocks. The corresponding analytical result from a derivation of 
Eq.~(\ref{inj_all}) is shown by the solid line in Fig.~\ref{diffuse}.
In both cases, the agreement is quite good.
\begin{figure}[t]
    \vspace{-3.2cm}
    \begin{center} 
      \epsfxsize9.0cm 
      \mbox{\epsffile{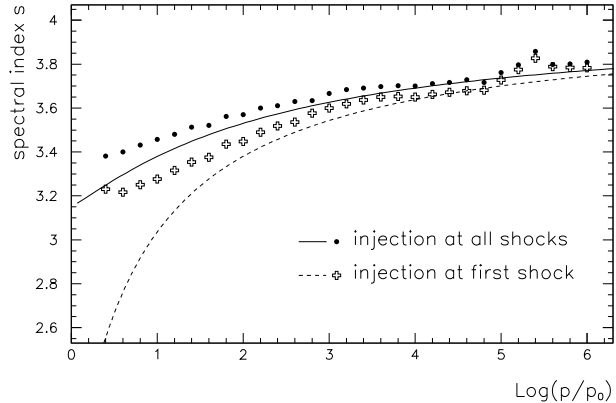}}
    \end{center}
    \vspace{-1.2cm} 
    \protect\caption{Discrete symbols show the Monte-Carlo results for the 
               spectral index $s$ vs. momentum $p$, downstream of 5th 
               quasi-parallel shock, including effects of escape with
               $\pesc =0.2$ and re-thermalisation. 
               Shock parameters are the 
               same as in Fig.~\ref{diffuse}. The lines are identical with 
               those in Fig.~\ref{diffuse} and are included here for 
               comparison.}
    \label{vgl_esc}
\end{figure}

In Fig.~\ref{vgl_esc} we have included effects of finite extent of the 
acceleration region by choosing the free parameter $\pesc=0.2$, as described 
above.
This leads essentially to a reduction of the number of shocks at which the
particles in the final downstream distribution have been accelerated,
and therefore the momentum distribution is steeper. 
In addition we remove particles from the the acceleration mechanism with
momentum $p<p_0$, which correspond to re-thermalisation. This produces
a cutoff of the test-particle distribution at $\log(p/p_0)=0$.
To show the effect of the escape, we have included in Fig.~\ref{vgl_esc}
the analytical results shown in Fig.~\ref{diffuse} as lines. 
The discrete symbols
represent the Monte-Carlo results. The number of initial particles here is
$2.1\cdot 10^6$. The symbols and the fit procedure used
are the same as described above. Note that the distribution of particles
injected at the first shock is a subset of the particles
injected at all shocks.  We do not perform independent runs of the code to 
measure these two momentum distributions. With increasing momentum,
the distribution of particles injected at all shocks has an increasing
fraction of particles which were injected at the first shock, because these 
have the highest probability to gain momentum. Therefore, the statistical
fluctuations of the Monte-Carlo distributions shown in Fig.~\ref{vgl_esc} 
(and also in Fig.~\ref{diffuse} and Fig.~\ref{obl}) are correlated at 
high momentum.
\subsection{Multiple oblique shocks}
\label{oblique}
By increasing the inclination angle $\Phi$, the ratio of the particle velocity
to the intersection velocity of magnetic field and shock becomes larger,
and the acceleration can no longer be described by the diffusion
approximation. Then, the calculations of Sect.~\ref{analytical} break down.

We consider here strong shocks with
speed $u_{\rm s}=0.1$, and $\Phi=70.53$.
The resulting velocity of the intersection point is
$u_{\rm s}/\cos\Phi=0.3$.
Such a shock produces a spectrum with canonical index $s_{\rm c}=3.45$
(Kirk \& Heavens~\cite{KiHe89}).
The realisation of the multiplicity we investigate here is one in which
the compression by the shock and the decompression of the plasma 
described in Sect.~\ref{analytical} are both effective only along the shock 
normal (which has the same direction for all subsequent shocks). 
Therefore, the decompression will restore the magnetic field to the initial 
upstream value and orientation. This realisation can be described by multiple
{\em identical} oblique shocks of the same inclination angle $\Phi$. 
This situation reveals most clearly the direct effect of the shock 
multiplicity, 
together with re-thermalisation and escape (for which we chose $\pesc=0.2$).
\begin{figure}[t]
    \vspace{-3.2cm}
    \begin{center} 
      \epsfxsize9.0cm 
      \mbox{\epsffile{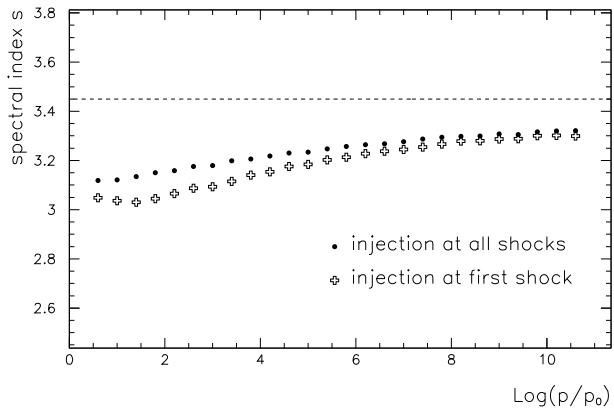}}
    \end{center}
    \vspace{-1.2cm} 
    \protect\caption{Discrete symbols show the Monte-Carlo results for the
                   spectral index $s$ vs. momentum $p$, downstream of 5th 
                   oblique shock. The dashed line indicates the canonical
	           spectral index $\sca=3.45$.
                  Parameters for all shocks are $r=4$, $u_{\rm s}=0.1$,
                  $\Phi=70.53$,
                   $u_{\rm s}/\cos\Phi=0.3$. With escape $\pesc =0.2$
                   and re-thermalisation included.}
    \label{obl}
\end{figure}
The results are presented in Fig.~\ref{obl}, where the discrete symbols 
represent the Monte-Carlo calculations in the same way as described above,
except that the linear interpolation of two adjacent bins is used up to
$\log(p/p_0)=5$, and an initial number of $5.4\cdot 10^6$ independent 
particles is simulated. 
The dashed line indicates the canonical spectral index $\sca$ for a single 
shock of this type. Although the escape between shocks leads to a steepening of
the spectrum towards the canonical result, the slope at multiple oblique shocks
is still flatter than for a single shock.
The {\em relative} hardening $\Delta/\sca\equiv(\sca-s)/\sca$ due to the 
multiplicity of the shock is in the high energy part (from about three
orders of magnitude above the injection momentum) quite similar 
to the diffusive case, shown in Fig.~\ref{vgl_esc}.
However, the main effect
of producing a hard spectrum is due to the obliquity
($\sca=3.45$ compared to $\sca=4$), and this is independent of momentum  
for relativistic particle velocities, as long no loss mechanisms are 
included. 
\section{Conclusion}
\label{conclusion}
Our Monte-Carlo simulation of first order Fermi acceleration, using the
shock-drift process and diffusively (or statistically) particle transport 
reproduced the analytical steady state test-particle spectrum at multiple 
shocks, for which the diffusion approximation does apply. 
We introduced a phenomenological re-thermalisation effect, which 
accounts for the fact, that below the injection momentum
no acceleration can occur. Furthermore,
with regard to the finite extension of the shock,
we allowed for escape of particles between subsequent shocks. 
This reduces the ability of multiple shocks to flatten the
spectrum which would be produced at one single shock.

As a more realistic setup of multiple shock acceleration at fast shocks
we considered oblique shocks, and included a finite escape probability 
$\pesc=0.2$ between shocks. 
We found very hard spectral distributions at strong shocks
($r=4$) with velocity $u_{\rm s}=0.1$ and obliquity $\Phi\simeq 70$.
The main effect here is produced by the obliquity, because the canonical 
spectral index is $s_{\rm c}=3.45$. 
In addition, the spectrum becomes even harder
with $s=3.21\pm 0.01$ at $\log(p/p_0)=4$ at the relatively low 
number $N=5$ of these moderately oblique shocks considered here 
(Fig.~\ref{obl}).

An example geometrical situation where multiple oblique shocks with (about) the
same inclination angle $\Phi$ are likely is a jet with helical magnetic 
field and along which a number of shocks exist. We assume, that the shocks
are propagating along the jet axis and do not modify the jet geometry.
The compression by the shock and the decompression of the plasma described 
in Sect.~\ref{analytical} are then both effective only along the shock normal. 
Therefore, the decompression will restore the initial upstream magnetic field
orientation. This situation can be described by multiple
{\em identical} oblique shocks. If the magnetic field would not be restored
to the upstream direction, the obliquity for subsequent plan-parallel 
shocks would increase, leading to even harder spectra than described above.

In regions, where 
geometrically uncorrelated shocks exist, like central regions of AGN,
a {\em large} number of {\em subsequent} shocks, through which a single 
distribution is processed, might be a too strong idealisation. 
Here, we would have to average over a low number of shocks with
different inclination angles $\Phi$. 
If we consider electrons, in addition effects of losses and magnetic 
field strength would generally have to be included.
However, our results indicate, that very flat synchrotron
spectra with index $0< \alpha=(s-3)/2 < 0.5$ can be produced
even at a low number ($N<10$) of oblique shocks.
\begin{acknowledgements}
U.G. acknowledges useful discussions with John Kirk. We thank the referee
for helpful comments on the manuscript.
This work was supported by the University of Minnesota Supercomputing 
Institute, by NSF grant AST-9619438 and by NASA grant NAG5-5055.
\end{acknowledgements}

\end{document}